\documentstyle[12pt]{article}
\begin{document}
\rightline{RU9305}
\vspace*{.5in}
\begin{center}
Left-Right Asymmetry in $\bar{p}p \rightarrow \pi^- \pi^+$\\

\vspace*{.8in}

George Bathas$^{\rm a}$ and W.M. Kloet$^{\rm a,b}$\\

\vspace*{.5in}

a. Department of Physics and Astronomy, Rutgers University\\
{}~~~~~P.O. Box 849 Piscataway NJ 08855-0849 USA\\

\vspace*{.3in}

b. Division de Physique Th\'{e}orique, Institut de Physique
Nucl\'{e}aire,\\
{}~~~~~   F-91406, Orsay Cedex, France
\end{center}

\vspace{1.2in}

\centerline{Abstract}

\bigskip

\indent Results for d$\sigma$/d$\Omega$ and $A_N$ in the reaction
$\bar{p}p \rightarrow \pi^-  \pi^+$
are predicted by a
simple quark model. They are compared to recent experimental data
from LEAR, as well as to previous predictions from nucleon-exchange models.
At low energy the quark model does better than the nucleon-exchange models,
but the overall comparison to experiment remains poor. In particular, the
double-dip structure of the experimental $A_N$ data is only partly represented.
This shortcoming of the simple quark model is traced back to a too small J=2
amplitude. This has interesting implications for the range of this specific
annihilation process.

\bigskip


\newpage

\baselineskip=18pt

\noindent  1. Introduction\\

Large left-right asymmetries in the reaction $\bar{p}p \rightarrow
\pi^- \pi^+$ are observed [1-5]
when the target proton is polarized perpendicularly to the scattering plane.
For example, when the spin of the target proton is up, the $\pi^-$
particles prefer
to emerge to the right of the $\bar{p}$ beam-direction at low initial
antiproton
momenta (e.g. p$_{\rm LAB}$ = 500 MeV/c, or T$_{\rm LAB}$ = 124 MeV),
while at high momenta (e.g. p$_{\rm LAB}$ = 1500 MeV/c, or T$_{\rm
LAB}$ = 832 MeV) the $\pi^-$ particles prefer to emerge to the
left.

     The reaction $\bar{p}p \rightarrow \pi^- \pi^+$ is one of the
simplest $\bar{p}p$ annihilation channels
because it depends on only two helicity amplitudes. In terms of partial waves
it depends only on the spin-triplet coupled waves, while isospin takes the
values I=0 for even J and I=1 for odd J. Therefore the existence of accurate
data, in particular for the analyzing power $A_N$ (also called $A_y$)
is a challenge to microscopic models (like the quark model) for the
$\bar{p}p$ annihilation into that
particular channel.

For a detailed review of $\bar{p}p$ annihilation in general we refer
to ref.[6].
Previous work that explicitly compares theoretical predictions for the above
reaction $\bar{p}p \rightarrow \pi^- \pi^+$ with the available
d$\sigma$/d$\Omega$  and $A_N$ data is based on an
annihilation mechanism driven by nucleon- and delta-exchange [5,7,8].
Reasonable agreement with the experimental structure of $A_N$  is obtained at
antiproton momenta in the range of 650-1100 MeV/c (T$_{\rm LAB}$ =
203-575 MeV). The
agreement improves towards the higher end of this momentum range. However, at
lower momenta (p$_{\rm LAB}$ = 467-585 MeV/c, T$_{\rm LAB}$ = 110-167
MeV), so far no model has
predicted an asymmetry of overall negative sign at all angles, as is observed
experimentally (see ref.[4,5]). Some reservation exists about
nucleon-exchange models. Certainly, a legitimate concern of models
based on nucleon-exchange is the distant extrapolation of the $\pi$NN
form factor, from on-shell
nucleons, into a region where one of the nucleons has become spacelike.
Typically in this extrapolation an effective cutoff-mass of 1.5-1.7 GeV has
been used.

It is therefore of interest to investigate an alternative
approach to a
microscopic annihilation mechanism, which is based on the constituent quark
model. In a simple quark model one does not specify which object is being
exchanged between the anti-proton and the proton, but rather how the three
quarks and three anti-quarks in the initial state can be rearranged and partly
annihilated in order to form two $q\bar{q}$ pairs, representing the
final two pions.

     In this paper we study the predictions at low energy for the differential
cross section d$\sigma$/d$\Omega$  and the analyzing power $A_N$  if the
annihilation mechanism
is described by a simple quark model. We compare the results of the quark
model with two other models where the annihilation mechanism is given by
nucleon-exchange. From comparing with the experimental data from refs.[4,5],
we are able to draw some conclusions about the specific annihilation
of $\bar{p}p$ into $\pi^- \pi^+$.

     In a more general context, one may also foster the hope that because of
the short range nature of the annihilation process the reaction
$\bar{p}p \rightarrow \pi^- \pi^+$
could be a testing ground for possible differences between a description in
terms of quarks and a description in terms of mesons, nucleons, and
isobars.\\

\noindent 2. Annihilation mechanism\\

     To describe the annihilation of anti-proton proton into two pions we
employ here the generalized $^3P_0$ and $^3S_1$ mechanism of ref.[9] .
The proton (and antiproton) wave function is described as a Gaussian

\begin{eqnarray}
\Psi_N (\vec{r}_1, \vec{r}_2, \vec{r}_3)~ =~ N_N~ {\rm exp}~[~ -
\frac{\alpha}{2} ~\Sigma~ (\vec{r}_{\rm i}~ - ~ \vec{r}_{\rm N})^2~]
{}~X_N ~{\rm (spin,~ isospin,~ color)},
\end{eqnarray}

\noindent where $\vec{r}_{\rm i}$ are the quark coordinates and
$\vec{r}_{\rm N}$ is the nucleon coordinate. If we
take the pion to be an S-wave meson, its wavefunction is given by

\begin{eqnarray}
\Phi_M~(\vec{r}_1, \vec{r}_4)~= ~N_M~{\rm exp}~[~ - \frac{\beta}{2}
{}~\Sigma~ (\vec{r}_{\rm i} ~-~ \vec{r}_{\rm M})^2 ~ X_M ~ {\rm (spin,~
isospin,~ color)},
\end{eqnarray}

\noindent where $\vec{r}_1$ and $\vec{r}_4$ are respectively the quark
and antiquark coordinates, and $\vec{r}_{\rm M}$
is the coordinate of the meson. Typical parameter values are $\alpha$
= 2.8 fm$^{-2}$ and
$\beta$ = 3.23 fm$^{-2}$, giving a nucleon radius of 0.60 fm and a
meson radius of 0.48 fm.

     The transition potential for this process, predicted by the simple quark
model, was worked out in ref.[9]. It turns out to be a non-local potential,
dependent on the relative $\bar{p}p$ coordinate $\vec{r}$, the
relative $\pi \pi$ coordinate $\vec{r}~'$, and
the Pauli spin operator $\vec{\sigma}$. If one allows for two
mechanisms, the $^3P_0$ and  $^3S_1$
mechanisms, to be responsible for the annihilation of one of the
$q\bar{q}$ pairs, the
potential can be written as

\begin{eqnarray}
V_{\rm ann}~= ~V(^3P_0) ~+~  \lambda ~V(^3S_1).
\end{eqnarray}

We repeat here the form of V($^3P_0$) derived in ref.[9]:

\begin{eqnarray}
V(^3P_0)(\vec{r}~',\vec{r})~ & \sim & \\ \nonumber
\{ A_V~  i\vec{\sigma}.\vec{r}~' ~ {\rm sinh}(C\vec{r}~'.\vec{r}) &~+~&
B_V~  i\vec{\sigma}.\vec{r} ~ {\rm cosh}(C\vec{r}~'.\vec{r})\} ~ {\rm
exp}(A~r'^2 ~+~ B~ r^2),
\end{eqnarray}

\begin{eqnarray}
{\rm where}~ A_V~ &=&~ \frac{\alpha ( \alpha ~+~ \beta)} {2 ( 4 \alpha
{}~+~ 3 \beta)}, \\
B_V~&=&~ \frac{3 (5 \alpha^2 ~+~ 8 \alpha \beta ~+~ 3 \beta^2)}{2 (4
\alpha ~+~ 3 \beta)}, \\
A ~&=&~ ~-~ \frac{\alpha (5 \alpha ~+~ 4 \beta)}{2 ( 4 \alpha ~+~ 3
\beta)}, \\
B ~&=&~ ~-~ \frac{3 ( 7 \alpha^2 ~+~ 18 \alpha \beta ~+~ 9
\beta^2)}{8( 4 \alpha ~+~ 3 \beta)}, \\
C ~&=&~ ~-~ \frac{3 \alpha ( \alpha ~+~ \beta)}{2 (4 \alpha ~+~ 3
\beta)}.
\end{eqnarray}

The expression for V($^3S_1$) consists of a longitudinal part and a transversal
part. The longitudinal part has the same general form as eq.(4), the
transversal part is obtained from eq.(4) by interchanging sinh and cosh. Both
expressions, with the appropriate coefficients are given in ref.[9]. The
parameter $\lambda$ in eq.(3) gives the relative strength of the two
mechanisms. The
range and the nature of the non-localities of V$_{\rm ann}$ is a
function of the size-parameters $\alpha$ and $\beta$ of the proton and
the pion
respectively, mentioned in
eqs.(1,2). The annihilation potential V$_{\rm ann}$ of eq.(3)
contributes to all total angular momenta J of the $\bar{p}p$ system.\\

\noindent 3. Observables and basic amplitudes\\

     As initial state we choose the $\bar{p}p$ wavefunction as given
by the 1982 Paris
$\bar{p}p$ model [10] including the Coulomb interaction. As final
state we choose non-interacting pions. Later, we will discuss the
implications of different cases
of initial state interactions and final state interactions, which may be
substantial [11], in a different paper.

      The observables, which have been measured in this reaction, are the
differential cross section d$\sigma$/d$\Omega$  and the analyzing
power $A_N$. Both observables
are expressed in the two helicity amplitudes F$_{\rm ++}$($\theta$)
and F$_{\rm +-}$($\theta$) which fully
describe the reaction $\bar{p}p \rightarrow \pi^- \pi^+$. The angle $\theta$
is the CM angle between the
outgoing $\pi^-$ and the incoming anti-proton. The observables considered are

\begin{eqnarray}
d\sigma/d\Omega ~= ~(~|F_{\rm ++}|^2 ~+~ |F_{\rm +-}|^2)/2 ,\\
A_N~ d\sigma/d\Omega~ =~ Im~ (F_{\rm ++}~ F_{\rm +-}^*).
\end{eqnarray}

     For completeness and later reference, we note the relation of the two
helicity amplitudes to the angular momentum amplitudes
$f_{\ell}^{\rm J}$ and the Legendre
polynomials $P_J({\rm cos}\theta)$  and their derivatives $P_J'({\rm
cos}\theta)$. The indices J and $\ell$ are
respectively the total angular momentum, and orbital angular momentum of the
$\bar{p}p$ system. The angular momentum of the $\pi^- \pi^+$ system is
$\ell_{\pi \pi}$ = J.  We have [12]

\begin{eqnarray}
F_{\rm ++}(\theta) ~ =~ \frac{1}{p} ~ \sum_{\rm J} ~
\sqrt{(2J+1)/2}~ ~(~\sqrt{J} ~ f_{J-1}^J  ~ - ~ \sqrt{(J+1)} ~
f_{J+1}^J) ~P_J ({\rm cos}\theta),
\end{eqnarray}

\noindent and

\begin{eqnarray}
F_{+-}(\theta) ~=~ \frac{1}{p} ~ \sum_{\rm J} ~ \sqrt{(2J+1)/2}~~
(\sqrt{1/J} ~ f_{J-1}^J ~+~ \sqrt{1/(J+1)} ~ f_{J+1}^J)~P_J' ({\rm cos}\theta).
\end{eqnarray}

     The angular momentum amplitudes $f_{\ell}^J$ are basically
two-dimensional integrals in $r'$ and $r$ of the type

\begin{eqnarray}
I_{iJ \ell}~ = ~ \int \int r'^2 dr' r^2dr \Phi_J^{\pi \pi}(r') ~V_{iJ
\ell} (r',r) ~ \Psi_{J \ell}^{\bar{p}p} (r).
\end{eqnarray}

Here V$_{iJ \ell}(r',r)$
originates from the angular momentum decomposition of the
transition potential V$_{\rm ann}(\vec{r}~',\vec{r},\vec{\sigma})$
and takes on two possible forms

\begin{eqnarray}
V_{1J \ell}(r',r) = r~ {\rm exp}~[ A~ r'^2 ~  +~ B~ r^2]~~  j_J^{\rm
mod}(C r' r),
\end{eqnarray}

\noindent and

\begin{eqnarray}
V_{2J \ell}(r',r)~ =~ r'~ {\rm exp}~ [ A~ r'^2 ~ +~ B~
r^2]~~j_{\ell}^{\rm mod}(C r' r).
\end{eqnarray}

The symbol $j_{\ell}^{\rm mod}$ in eqs.(15,16) stands for the modified
spherical Bessel
function of the first kind of order $\ell$. In eq. (14) above
$\Phi_J^{\pi \pi}$($r'$) and $\Psi_{J \ell}^{\bar{p}p}$
(r) are the final $\pi \pi$ and initial $\bar{p}p$ wavefunctions respectively.

     As expected the integral of the above eq.(14) probes the
$\bar{p}p$ wavefunction
and $\pi \pi$ wavefunction at rather short distances since the values
of A, B, and C
are typically A = -1.80 fm$^{-2}$, B = -5.59 fm$^{-2}$, and C = -1.21
fm$^{-2}$.  On the other
hand one needs to realize that, for example, the wave function
$\Psi_{J \ell}^{\bar{p}p}$ (r)
vanishes at short distances r. This is caused by the total $\bar{p}p$
annihilation
predominantly in channels different from $\pi \pi$, which, in case of
the Paris model
[10] is described by a phenomenological annihilation potential. This of course
means to some extent a double counting of the specific annihilation
into $\pi \pi$.
But since the annihilation into $\pi \pi$ only accounts for about a
percent of the
total annihilation, double counting due to use of the Paris model for the
initial $\bar{p}p$ interaction can be ignored. We can safely use this
method to study
the microscopic behaviour of the specific annihilation $\bar{p}p
\rightarrow \pi^- \pi^+$. From the
above it is clear that integrals of the type of eq.(14) are sensitive to the
model used for $\bar{p}p$ initial state interaction, and at the same
time sensitive to
the model for $\pi \pi$ final state interaction. We return to this
aspect in a separate paper.\\

\noindent 4. Results\\

If we take for $\Psi_{J \ell}^{\bar{p}p}$ (r) the Paris model of 1982
[10] and for $\Phi_J^{\pi \pi}$ ($r'$)
simply plane waves, (no $\pi \pi$ interaction as they go out) , we
obtain the result
at p$_{\rm LAB}$ = 497 MeV/c (T$_{\rm LAB}$ = 123 MeV) for
differential cross section d$\sigma$/d$\Omega$  and
analyzing power $A_N$, shown in fig.1 for the case $\lambda$ = - 0.5.
The parameter $\lambda$
which gives the relative strength of the $^3S_1$ versus $^3P_0$
mechanisms, can be
varied.
The other free parameter, the overall strength off V$_{\rm ann}$
is of no
consequence for $A_N$ while for d$\sigma$/d$\Omega$ it is chosen so as
to obtain the correct
experimental total cross section. The values for $\alpha$ and $\beta$
are kept fixed. With
$\lambda$ = -0.5 one is able to obtain an overall negative $A_N$ at
p$_{\rm LAB}$ = 497 MeV/c
(T$_{\rm LAB}$ = 123 MeV), although the forward minimum is not deep
enough. The relative
poorness of the prediction is however still better than the predictions at
similar energies by the nucleon-exchange models [5,7,8] as is shown in fig.2.
Both N-exchange models give a forward positive value of $A_N$ at low energy, as
long as the final $\pi \pi$ interaction is ignored. On the other hand,
when $\pi \pi$ FSI is
included in the N-exchange model [11], the forward $A_N$ turns negative but the
backward $A_N$ becomes positive instead. As mentioned above, experimental data
for $A_N$ at these energies are negative at all angles. By itself, it
is amazing
that this simple quark model is able to predict observables that show any
resemblance to experiment.

The number of partial waves needed at this momentum p$_{\rm LAB}$ is
rather small.
In fig.3 it is shown that already J$<$4 and J$<$5 cannot be
distinguished. This
means that at this energy J$<$5 is certainly sufficient.

The prediction at higher momentum, p$_{\rm LAB}$ = 679 MeV/c (T$_{\rm
LAB}$ = 210 MeV) with
no change in the parameters $\lambda$, $\alpha$, $\beta$, and the
overall strength of V$_{\rm ann}$, is
given in fig.4. Again the forward negative dip in $A_N$ is not deep
enough. It is
however encouraging that with the same value $\lambda$ = -0.5 the
quark model now
predicts a positive value for $A_N$ near cos$\theta$ = 0, while the
negative forward and
backward values for $A_N$ remain. The experimental data at this
energy[4,5] show
the same trend of a double dip with a positive hump at 90 degrees. It is still
sufficient at this energy, to include only amplitudes with J$<$5. For easy
comparison also the nucleon-exchange predictions of refs.[5,7,8] are shown in
fig.4.

      One particularly observes, in this low energy range, the general failure
of all models to predict the large forward dip in $A_N$, simultaneously with a
large negative value in the backward direction.\\

\noindent 5. Discussion of the results\\

     We note that predictions at low energy for d$\sigma$/d$\Omega$
and $A_N$ in the simple
quark model are better or of the same quality as for N-exchange models. Of
course we have a free parameter $\lambda$ to play with that gives the relative
strength of the $^3S_1$ and $^3P_0$ mechanisms, as well as an overall factor
that sets
the scale of the cross section.

     At this point one can ask the question, why can one not do better? To
answer that question, it is helpful to explicitly check the angular behaviour
of the terms in the sum over J in the amplitudes $F_{++}$ and $F_{+-}$
of eqs.(12,13) which are

\begin{eqnarray}
F_{++} (\theta)~ \sim ~ -~ \sqrt{1/2}~~~ ^3P_0 ~+~ \sqrt{3/2}~~ (~^3S_1
{}~-~ \sqrt{2}~~^3D_1)~ {\rm cos} \theta\\ \nonumber
{}~ +~ \sqrt{5/2}~  (~\sqrt{2}~ ^3P_2 ~-~ \sqrt{3}~ ^3F_2)~(3~ {\rm
cos}^2 \theta - 1)/2 + \ldots ~,
\end{eqnarray}

\noindent and

\begin{eqnarray}
F_{+-} (\theta)~ \sim ~  -~  \sqrt{3/2}~~(~ ^3S_1 ~+~ \sqrt{1/2}~ ~ ^3D_1)~
{\rm sin}\theta\\ \nonumber
{}~ -~  \sqrt{5/2}~~(\sqrt{1/2}~ ~ ^3P_2 ~+~  \sqrt{1/3}~ ~ ^3F_2)~(3~
{\rm sin} \theta {\rm cos} \theta)~ + ~ \ldots .
\end{eqnarray}

\noindent From inspection it is clear that for only J=0,1 the
asymmetry $A_N$ has the form

\begin{eqnarray}
A_N ~ \sim ~ {\rm sin} (\theta) ~ (a_0 ~+~ a_1 ~{\rm cos} ~ (\theta)).
\end{eqnarray}

\noindent At low energies, the expression of eq.(19) seems to be the
basic form of the
predicted $A_N$ in both N-exchange models, and to some extend the same
holds for
the quark model (see fig.1b). To get an angular behaviour of $A_N$ that is of
the
pronounced double-dip structure as in experiment, one needs at least a rather
substantial J=2 (e.g. $^3P_2 - ^3F_2$) contribution.

     This can be clearly demonstrated by a simple toy-model with only $^3P_0$,
$^3S_1$, and  $^3F_2$ $\bar{p}p$ amplitudes (replacing $^3F_2$ by only
$^3P_2$ does not work as well
due to their opposite signs in $F_{++}$ ($\theta$) in eq.(17)). In
fig.5  we show the
result of such a toy-model where $^3P_0$, $^3S_1$, and $^3F_2$ have
relative strength  1 : 0.8 : 0.5 respectively. The phases of $^3P_0$
and $^3F_2$ are the same, $^3S_1$ has a
phase difference of 3$\pi$/2 with the other two amplitudes. Here the
double-dip structure of the asymmetry $A_N$, shown in fig.5, is due to
a strong J=2 presence.

     Going back to the quark model, we observe that indeed there the J=2
amplitude is relatively small compared to the J=0,1 amplitudes, and that is
the main reason why $A_N$ has not the pronounced double-dip structure
as seen in experiment.\\

\newpage

\noindent 6. Conclusions\\

     One shortcoming of both quarkmodel and N-exchange models may be that the
transition potential V$_{\rm ann}$ ($\vec{r}~'$,$\vec{r}$,$\vec{s}$)
is of too short a range to make the J=2
amplitudes play a significant role at low energies. In the simple quark model
the range of V$_{\rm ann}$ is determined by the coefficients A,B, and C in the
exponent. Their values are set by the nucleon size parameter $\alpha$
and pion size
parameter $\beta$. In order to increase the range of the potential
(and enhance the role of the J=2 amplitude) one can decrease the
values of $\alpha$ and $\beta$.

     As an example we show in fig.6 the result at p$_{\rm LAB}$ = 497
MeV/c for $\alpha$ = 2.8 fm$^{-2}$ and lowering $\beta$ to 1.73
fm$^{-2}$ (dot-dashed curve). This would correspond with
a pion radius of 0.66 fm. The previous value was r$_{\pi}$ = 0.48 fm.
Indeed one sees
that the double-dip structure of $A_N$ is enhanced. At the same time the
differential cross section d$\sigma$/d$\Omega$  is somewhat more
forward peaked, in better
agreement with the experimental shape. In the case of
d$\sigma$/d$\Omega$  the overall
strength was adjusted to give the same value at cos $\theta$ = -0.5. The
corresponding values for A, B, and C are A = -1.79 fm$^{-2}$, B = -
3.87 fm$^{-2}$, and
C = - 1.16 fm$^{-2}$. Similar effects in the asymmetry $A_N$ can be obtained by
keeping $\beta$ fixed at $\beta$ = 3.23 and reducing $\alpha$. The
dashed curve in fig.6 shows
the result for $\alpha$ = 1.24 fm$^{-2}$ (r$_p$ = 0.90 fm ). In that
case the values for A,
B, and C become A = - 0.81 fm$^{-2}$, B = - 4.52 fm$^{-2}$, and C = -
0.57 fm$^{-2}$. For
the above two cases the value of $\lambda$ is - 0.45. In both cases
the role of the
J=2 amplitudes is enhanced, when compared with the model with the conventional
values for $\alpha$ and $\beta$ (solid curve).

     Overall, it seems that experiment requires at this energy a significant
J=2 amplitude. One way to accomplish this, is to increase the range of the
transition potential as shown above. Of course one can also search for the
reason of the discrepancy between theory and experiment, in the model
dependence of the initial and final state interaction. Work on this is
forthcoming.

     Finally, because the reaction $\bar{p}p \rightarrow \pi^{\circ}
\pi^{\circ}$
has only isospin I = 0 amplitudes
and therefore only even J contributions, the measurement of
$A_N$ for
that process, if ever performed, may shed additional light on the microscopic
process of $\bar{p}p$ annihilation into two mesons.\\

\newpage

\noindent Acknowledgment\\

     One of the authors (W.M.K) wishes to thank the Institut de Physique
Nucl\'{e}aire at Orsay for its warm hospitality. The authors are
grateful to B. Loiseau for stimulating discussions and for making
available the $\bar{p}p$ results of the Paris model.\\

\noindent References\\

\noindent [1] E.\ Eisenhandler et al, Nucl.Phys. B69, 109 (1975).\\
\noindent [2] A.\ A.\ Carter et al, Nucl.Phys. B127, 202 (1977).\\
\noindent [3] T.\ Tanimori et al, Phys.Rev.Lett. 55, 1835 (1985).\\
\noindent [4] R.\ Birsa et al, Nucl.Phys.(Proc.Suppl.) B8, 141 (1989).\\
\noindent [5] R.\ Birsa et al, Proc. 1st Biennial Conf. on Low Energy
Antiproton Physics (Stockholm 1990), ed P.\ Carlson, A.\  Kerek, S.\ Szlagyi
(World Scientific, Singapore, 1991) p. 180.\\
\noindent [6]  C.\ B.\ Dover, T.\ Gutsche, M.\ Maruyama, A.\ Faessler, Prog.
in Particle and Nuclear Physics 29, 87 (1992).\\
\noindent [7] B.\ Moussallam, Nucl. Phys. A407, 413 (1983); A429, 429 (1984).\\
\noindent [8] V.\ Mull, J.\ Haidenbauer, T.\ Hippchen, K.\ Holinde,
Phys. Rev. C44, 1337 (1991).\\
\noindent [9] G.\ Bathas, W.\ Kloet, Phys. Lett. B301, 155 (1993).\\
\noindent [10] J.\ C\^{o}t\'{e}, M.\ Lacombe, B.\ Loiseau, B.\ Moussallam, R.\
Vinh Mau, Phys. Rev. Lett. 48, 1319 (1982).\\
\noindent [11] V.\ Mull, K.\ Holinde, J.\ Speth, Phys. Lett. B275, 12
(1992).\\
\noindent [12] A.\ D.\ Martin, M.\ R.\ Pennington, Nucl. Phys. B169, 216
(1980).\\

\newpage

\noindent FIGURE  CAPTIONS\\

\noindent FIG. 1. Differential cross section and analyzing power at
p$_{\rm LAB}$ = 497 MeV/c.
Parameter values are $\lambda$ = - 0.5, $\alpha$ = 2.80 fm$^{-2}$, and
$\beta$ = 3.23 fm$^{-2}$. Data for
d$\sigma$/d$\Omega$  are from ref.[3], and $A_N$ data are from ref.[5].\\

\noindent FIG. 2. Analyzing power at p$_{\rm LAB}$ = 497 MeV/c. The solid curve
is the quark
model. The nucleon exchange model is represented by the dashed line
ref.[5,7], and the dotdash line from ref.(8). All models are without
$\pi \pi$ FSI.
Data are from ref.[5].\\

\noindent FIG. 3. Effect of various partial waves at p$_{\rm LAB}$ = 497
MeV/c. The solid curve is
the prediction for J$<$5, and cannot be distinguished from J$<$4. the
dashed curve is J$<$3, and the dotdash curve is J$<$2.\\

\noindent FIG. 4. Differential cross section and analyzing power at
p$_{\rm LAB}$ = 679 MeV/c. The
solid curve is the quark model, with the same parameters as in fig.1. The
dashed curve is ref.[5,7], and the dotdash curve is ref.[8]. Data for
d$\sigma$/d$\Omega$
are from ref.[3], and $A_N$ data are from ref.[5].\\

\noindent FIG. 5. Result from simple toy-model with $^3P_0$, $^3S_1$,
and $^3F_2$ amplitudes. For
details see text.\\

\noindent FIG. 6. Results at p$_{\rm LAB}$ = 497 MeV/c for $\alpha$ = 2.8
fm$^{-2}$  and $\beta$ = 1.73 fm$^{-2}$ (dotdash curve). Dashed curve
is for $\alpha$ = 1.24 fm$^{-2}$ and $\beta$ = 3.23 fm$^{-2}$. In these first
two curves $\lambda$ = -0.45. The solid curve serves as reference to
the earlier values $\alpha$ = 2.8 fm$^{-2}$ and $\beta$ = 3.23
fm$^{-2}$, where $\lambda$ = -0.50. Data are from refs.[3,5].

\end{document}